\documentclass[aps,pra,twocolumn,floatfix,twoside]{revtex4}

\usepackage{amssymb}
\usepackage{graphicx}
\usepackage{dcolumn}
\usepackage{bm}

\begin{document}

\title{Condensate fraction of molecules for a spin mixture of ultracold
fermionic atoms}
\author{Hongwei Xiong$^{\ast}$}
\affiliation{State Key Laboratory of Magnetic Resonance and Atomic and Molecular Physics,
Wuhan Institute of Physics and Mathematics, Chinese Academy of Sciences,
Wuhan 430071, P. R. China}
\author{Shujuan Liu}
\affiliation{State Key Laboratory of Magnetic Resonance and Atomic and Molecular Physics,
Wuhan Institute of Physics and Mathematics, Chinese Academy of Sciences,
Wuhan 430071, P. R. China}
\author{Min Liu}
\affiliation{State Key Laboratory of Magnetic Resonance and Atomic and Molecular Physics,
Wuhan Institute of Physics and Mathematics, Chinese Academy of Sciences,
Wuhan 430071, P. R. China}
\author{Kelin Kao}
\affiliation{State Key Laboratory of Magnetic Resonance and Atomic and Molecular Physics,
Wuhan Institute of Physics and Mathematics, Chinese Academy of Sciences,
Wuhan 430071, P. R. China}
\author{Mingsheng Zhan}
\affiliation{State Key Laboratory of Magnetic Resonance and Atomic and Molecular Physics,
Wuhan Institute of Physics and Mathematics, Chinese Academy of Sciences,
Wuhan 430071, P. R. China}
\date{\today }

\begin{abstract}
The condensate fraction of molecules for ultracold Fermi gases is
investigated for the magnetic field below the Feshbach resonant magnetic
field. Assuming that there is no loss of particles and energy during the
adiabatic magnetic-field sweep, a simple theory is used to interpret the
measured condensate fraction in the experiments by JILA group (Phys. Rev.
Lett. 92, 040403 (2004)) and MIT group (Phys. Rev. Lett. 92, 120403 (2004)).
Our theory shows that the condensate fraction of molecules is dependent on
the initial condition of the system and especially on the process of the
magnetic-field sweep.


$\ast$ Electronic address: xionghongwei@wipm.ac.cn
\end{abstract}

\maketitle

The evidence for Bose-Einstein condensates of diatomic molecules has been
finally given in several remarkable experiments \cite{JOCHIM,JIN,MIT-mole}
which will obviously lead to intensive theoretical and experimental
researches on the ultracold Fermi gases. The magnetic-field Feshbach
resonance \cite{FESHBACH-THE,FECH-EXP} plays an important role in most of
the recent experiments on ultracold Fermi gases because it can change both
the strength and sign of the scattering length $a$ between fermionic atoms
with different internal freedom. On the side of strongly repulsive
interaction (BEC side), there is molecule which is short-range fermionic
pairs. On the side of strongly attractive interaction (BCS side), one
expects that there are fermionic pairs analogously to the electronic Cooper
pairs in superconductor. The Feshbach resonance has given us an important
opportunity to investigate the pairing phenomena in the ultracold Fermi
gases especially the BCS-BEC crossover which has been discussed in a lot of
theoretical researches \cite%
{LEGGETT,NOZI,STOOF,TIMM,OHASHI1,MILSTEIN,STAJ,CARR,BRUUN,FALCO,XIONGCROSS,XIONGON}%
. Recently, the BCS-BEC crossover is investigated by several experiments
\cite{JIN-fermion,MIT-fermion,GRIMM,SALOMON,THOMAS,GRIMM1,GRIMM2} such as
the superfluid hydrodynamics \cite{THOMAS,GRIMM1} and pairing gap \cite%
{GRIMM2}.

The condensate fraction near a Feshbach resonance has been investigated
experimentally in \cite{JIN-fermion,MIT-fermion} by adiabatically sweeping
the magnetic field to different value, and the condensate fraction is then
measured by a following fast decreasing of the magnetic-field to convert the
dimers of fermionic atom pairs into bound molecules. Above the Feshbach
resonant magnetic field, the condensate fraction observed by JILA group \cite%
{JIN-fermion} is investigated in \cite{FALCO} based on the model of a
molecular BEC. On the Feshbach resonance, a simple theoretical model of a
mixture of Fermi degenerate gas and dimeric gas \cite{XIONGON} is proposed
to explain the high condensate fraction observed in \cite{MIT-fermion}.
Below the Feshbach resonant magnetic field, the condensate fraction of
molecules is investigated experimentally by JILA group \cite{JIN-fermion}
for $^{40}K$ and MIT group \cite{MIT-fermion} for $^{6}Li$ with different
magnetic-field sweep process. As far as we know, below the Feshbach resonant
magnetic field, presently there is no theoretical interpretation for the
condensate fraction of molecules measured in \cite{JIN-fermion,MIT-fermion}.
In the experimental work of JILA group \cite{JIN-fermion}, the condensed
atom pairs was found to be at most $15\%$, while the MIT group \cite%
{MIT-fermion} observed a much higher fraction of condensed atom pairs below
the Feshbach resonant magnetic field. The high condensate fraction observed
by MIT group has been confirmed by other group such as the experiment by
Bartenstein et al \cite{GRIMM} for $^{6}Li$. Thus, there is an urgent need
for a theoretical model to interpret this great difference for different
groups. In the present work, omitting the loss of particles and thus the
loss of energy during the adiabatic magnetic-field sweep, a very simple
theoretical model is used to interpret the experimental results in both \cite%
{JIN-fermion} for $^{40}K$ and \cite{MIT-fermion} for $^{6}Li$. It is found
that the condensate fraction of molecules is dependent on the initial
condition of the system and especially on the process of the magnetic-field
sweep.

The magnetic-field Feshbach resonance \cite{FESHBACH-THE,FECH-EXP} can
change both the strength and sign of the atomic interaction. Near the
Feshbach resonance, the scattering length is $a\left( B\right) =a_{bg}\left[
1-w/\left( B-B_{0}\right) \right] $ with $a_{bg}$ being the background
scattering length and $B$ denoting the magnetic field with Feshbach resonant
magnetic field $B_{0}$. Below the Feshbach resonance, the energy of the
molecular state\textit{\ }can be estimated well to be $-\hbar ^{2}/ma^{2}$
\cite{STOOF-REPORT} with $m$ the mass of the fermionic atom.

During an evaporative cooling process, it is well-known that there is a
significant loss of energy and particles. For most of the recent
experiments, however, the BCS-BEC crossover is investigated for the
ultracold gases prepared at an initial magnetic field after different
evaporative cooling process. After the evaporative cooling and during the
following adiabatic magnetic-field sweep, the loss of particles can be
omitted and thus we omit the loss of the energy of the system. In the
present experiments on the BCS-BEC crossover, the temperature is much
smaller than the Fermi temperature. Thus, in this work, we investigate the
ultracold gases at zero temperature. At zero temperature and during an
adiabatic magnetic-field sweep process, omitting the loss of particles and
thus the loss of energy, in the present work a simple theory is used to
investigate the condensate fraction during the adiabatic magnetic-field
sweep.

Assume that $N$ denotes the total number of fermionic atoms in the absence
of molecules and $N_{m}$ is the total number of molecules. Assume further
that $N_{\uparrow }$ and $N_{\downarrow }$ are the number of fermionic atoms
with different internal freedom. In the present experimental works on the
BCS-BEC crossover, an equal mixture of fermionic atoms with different
internal freedom is prepared in an optical trap. In this case, one has the
following confinement condition:

\begin{equation}
N=2N_{\uparrow }+N_{m}.  \label{atomconservation}
\end{equation}

To use the assumption that the loss of energy can be omitted, we calculate
the overall energy of the system below and on the Feshbach resonant magnetic
field. Below the Feshbach resonant magnetic field with $\left\vert
a\right\vert /\overline{l}<1$ ($\overline{l}$ is the average distance
between atoms), the chemical potential of the Fermi gas is given by

\begin{equation}
\mu _{\uparrow }\left( N_{\uparrow },B\right) =\frac{\left( 6\pi ^{2}\right)
^{2/3}\hbar ^{2}}{2m}n_{\uparrow }^{2/3}+\frac{1}{2}g_{F}n_{\uparrow }+V_{F},
\label{chemicalbelow}
\end{equation}%
where the coupling constant $g_{F}=4\pi \hbar ^{2}a/m$, and $n_{\uparrow }$
is the density distribution of the fermionic atoms in an internal freedom. $%
V_{F}=m(\omega _{x}^{2}x^{2}+\omega _{y}^{2}y^{2}+\omega _{z}^{2}z^{2})/2$
is the external potential of the fermionic atom. The overall energy $%
E_{F}\left( N_{\uparrow },N_{\downarrow }\,,B\right) $ of the interacting
Fermi gases is given by

\begin{equation}
E_{F}\left( N_{\uparrow },N_{\downarrow },\,B\right) =\frac{\left(
6N_{\uparrow }\right) ^{4/3}\hbar \omega _{ho}}{4}+g_{F}\int n_{\uparrow
}^{2}dV,
\end{equation}%
where $\omega _{ho}=\left( \omega _{z}\omega _{y}\omega _{z}\right) ^{1/3}$
is the geometric average of the harmonic frequencies.

At zero temperature, by using the Thomas-Fermi approximation \cite{RMP}, the
overall energy of $N_{m}$ molecules in the condensate is given by

\begin{equation}
E_{m}\left( N_{m},B\right) =N_{m}\varepsilon _{m}+\frac{5N_{m}\mu _{m}^{0}}{7%
},  \label{energy-bec}
\end{equation}%
where the energy of the molecular state is $\varepsilon _{m}=-\hbar
^{2}/ma^{2}$. In the above expression, $\mu _{m}^{0}=\hbar \omega
_{ho}\left( 15N_{m}a_{m}/a_{ho}^{m}\right) ^{2/5}/2$ with the molecular
scattering length $a_{m}\approx 0.6a$ \cite{PETROV} and the harmonic
oscillator length $a_{ho}^{m}=\sqrt{\hbar /2m\omega _{ho}}$. The overall
energy of the mixture gases of fermionic atoms and molecules is then%
\begin{equation}
E_{F-m}=E_{F}\left( N_{\uparrow },N_{\downarrow }\,,B\right) +E_{m}\left(
N_{m},B\right) .
\end{equation}

On resonance, the absolute value of the scattering length $a$ is divergent
and thus much larger than the average distance $\overline{l}$ between
particles. In this situation, the ultracold gas can be described in the
unitarity limit \cite{HEISELBERG,CARLSON,HO1,HO} where there is a universal
behavior for the system. By using the local density approximation, the
general form of the chemical potential for the Fermi gas in the unitarity
limit can be given by the dimensionality analysis:

\begin{equation}
\mu _{\uparrow }^{res}=\left( 1+\beta \right) \frac{\left( 6\pi ^{2}\right)
^{2/3}\hbar ^{2}}{2m}n_{\uparrow }^{2/3}+V_{F},  \label{chemical}
\end{equation}%
where $\beta $ is the correction due to the strong interaction between
fermionic atoms with different internal freedom on resonance. The parameter $%
\beta $ has been investigated by both experiments \cite%
{GRIMM,SALOMON,THOMAS1} and theories \cite{HEISELBERG,CARLSON}. On
resonance, a quantum Monte Carlo calculation \cite{CARLSON} gives $\beta
=-0.56$. From Eq. (\ref{chemical}), the overall energy of the system on
resonance is given by

\begin{equation}
E_{res}\left( N\right) =\left( 1+\beta \right) ^{1/2}\frac{\left( 3N\right)
^{4/3}}{4}\hbar \omega _{ho}.  \label{energy-res}
\end{equation}

We turn to discuss the condensate fraction during the adiabatic
magnetic-field sweep with different initial condition of the system. Firstly
we consider the case that the initial system in thermal equilibrium is
prepared on resonance, and then the magnetic field is lowered adiabatically
below the Feshbach resonant magnetic field $B_{0}$. In this situation, if
the loss of the energy during the magnetic-field sweep is omitted, one has
the following confinement condition on the condensate fraction

\begin{equation}
E_{F}\left( N_{\uparrow },N_{\downarrow }\,,B\right) +E_{m}\left(
N_{m},B\right) =E_{res}\left( N\right) .  \label{energy-cons}
\end{equation}%
Combining with Eq. (\ref{atomconservation}), one can get the condensate
fraction of molecules $x=2N_{m}/N$ at zero temperature for the magnetic
field below $B_{0}$.

In the experiment by JILA group \cite{JIN-fermion} for $^{40}K$, the
scattering length is $a=174a_{0}\left( 1-7.8/\left( B-B_{0}\right) \right) $
with $a_{0}$ being the Bohr radius and $B_{0}=202.1$\textrm{G }being the
Feshbach resonant magnetic field. In addition, according to the experiment
in \cite{JIN-fermion}, $N\approx 2\times 10^{5}$ and $\omega _{x}/2\pi
=\omega _{y}/2\pi =320$ \textrm{Hz}, $\omega _{z}=\omega _{x}/79$. In \cite%
{JIN-fermion}, the system was initially prepared at a magnetic field $235.6$
\textrm{G} far above the Feshbach resonance. The magnetic field was then
swept across $B_{0}$\ adiabatically and especially there was a holding time
of about $3$ \textrm{ms} on resonance (see the inset of figure 2 in \cite%
{JIN-fermion}). The adiabaticity of this sweep process can be verified by
estimating the relaxation time $\gamma \approx 1/\sqrt{2}n_{\uparrow }\sigma
\overline{v}$ with the total cross section $\sigma =8\pi a^{2}$ and the mean
value of the velocity of the thermal atoms $\overline{v}=\sqrt{8k_{B}T/\pi m}
$. A simple calculation shows that for the regime of the magnetic field
investigated in \cite{JIN-fermion}, the relaxation time is smaller than $0.1$
\textrm{ms}, while the sweep and holding time in \cite{JIN-fermion} is
larger than $1$ \textrm{ms}. One should note that the condensate fraction
was measured in \cite{JIN-fermion} by lowering quickly (non-adiabatically)
the magnetic field which results in a measured loss of $50\%$ of the
molecules \cite{JIN-fermion}.

In \cite{JIN-fermion}, there was a holding time of about $3$ \textrm{ms} on
resonance. On the other hand, the divergent scattering length on resonance
means that the relaxation time is extremely small and a small loss of
particles on resonance will change the overall energy of the system
effectively. Thus the energy of the ultracold gases on resonance can be
determined by $E_{res}\left( N\right) $. In this situation, Eqs. (\ref%
{atomconservation}) and (\ref{energy-cons}) can be used to calculate the
condensate fraction $x$ for different magnetic field $B$ ($<B_{0}$). The
dotted line in figure 1 shows the condensate fraction calculated from Eq. (%
\ref{energy-cons}) by using the above experimental data. We see that at
least qualitatively the condensate fraction based on $\beta =-0.56$ agrees
with the experimental result shown by the solid circle. The solid line shows
the condensate fraction when $50\%$ loss of the molecules is considered
during the probing procedure (i.e. lowering the magnetic field quickly)
pointed out in \cite{JIN-fermion}.

\begin{figure}[tbp]
\includegraphics[width=0.8\linewidth,angle=270]{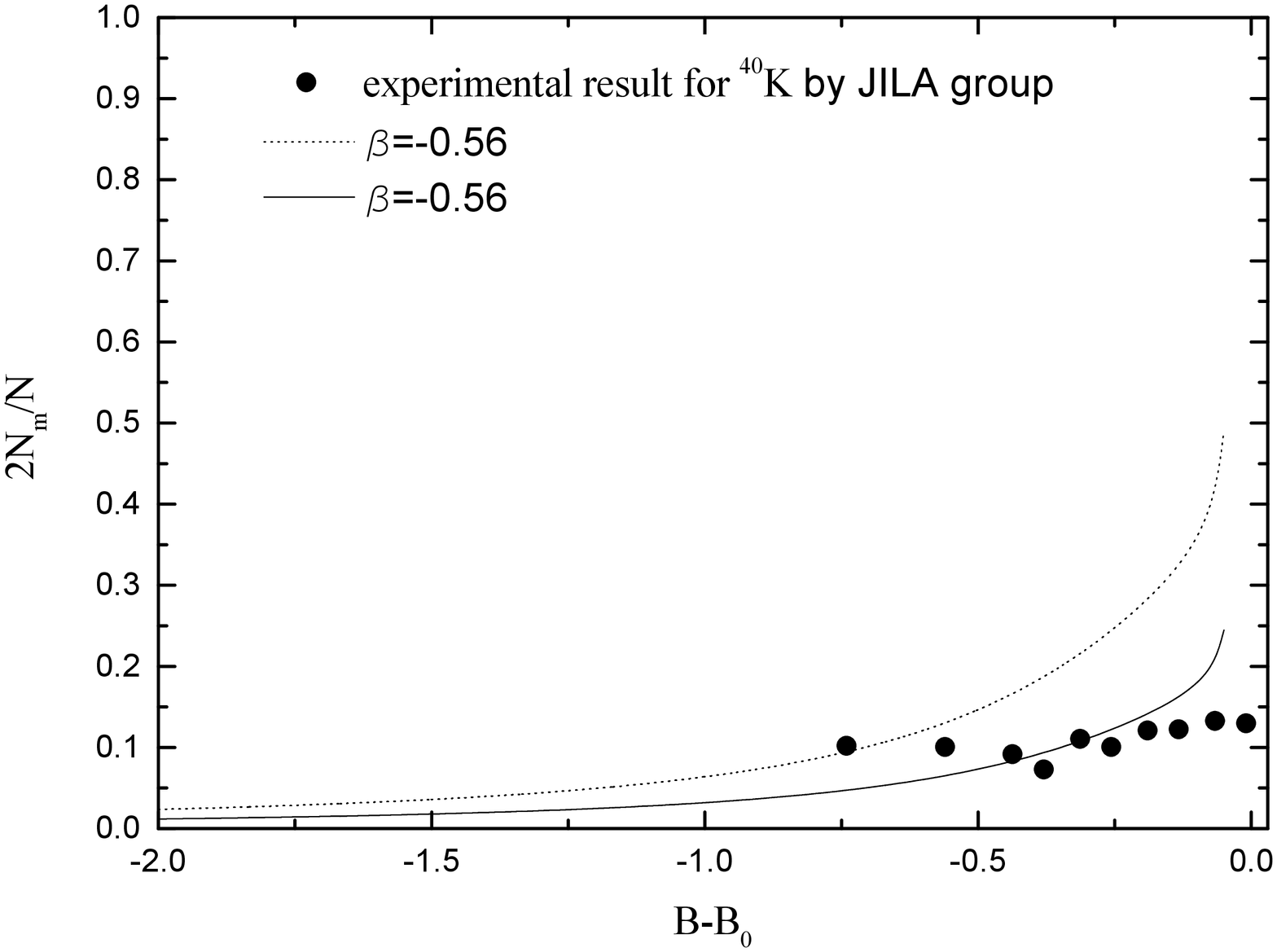}
\caption{By using the experimental parameters of JILA group \cite{JIN-fermion}%
, shown is the condensate fraction $x=2N_{m}/N$ for $^{40}K_{2}$
molecules at zero temperature and below the Feshbach resonant
magnetic field $B_{0}$. The parameter $\beta =-0.56$ is the
correction to the energy of the ultracold gases on resonance. The
dotted line is the condensate fraction obtained from Eq.
(\ref{energy-cons}), while the solid line is the condensate
fraction obtained by considering the loss of molecules during the
probing procedure. The solid circles are the measured condensate
fraction by JILA group.}
\end{figure}

We now consider another case that the system is prepared initially below the
Feshbach resonant magnetic field (i.e. on the side of molecular BEC).
Assuming that $x_{ini}$ is the condensate fraction at an initial magnetic
field $B_{ini}$, the condensate fraction after lowering adiabatically the
magnetic field to a value $B$ is then determined by the following
confinement condition when the loss of energy is omitted:

\[
{\ E_{F}\left( N_{\uparrow }=N_{\downarrow }=N\left( 1-x_{ini}\right)
/2,B_{ini}\right) +E_{m}\left( x_{ini}N/2,B_{ini}\right) }
\]

\begin{equation}
={E_{F}\left( N_{\uparrow }=N_{\downarrow }=N\left( 1-x\right) /2,B\right)
+E_{m}\left( Nx/2,B\right) }.  \label{Ketterle}
\end{equation}

Different from the experiment by JILA group \cite{JIN-fermion}, the
experimental results illustrated by $\square $ in figure 3 of the
experimental paper by MIT group \cite{MIT-fermion} were observed for the
system initially prepared at the magnetic field $B_{ini}=770$ \textrm{G }%
\cite{MIT-fermion} which is smaller than $B_{0}$. From the experimental
data, the condensate fraction at $B_{ini}=770$ \textrm{G }is $x_{ini}=0.56$.
The condensate fraction of molecules below $B_{0}$ was then investigated
experimentally by lowering adiabatically the magnetic field. One can easily
verify the adiabaticity of the magnetic-field sweep in \cite{MIT-fermion}
through the estimation of the relaxation time. The experimental parameters
in \cite{MIT-fermion} are $a=-1018a_{0}\left( 1+325/\left( B-B_{0}\right)
\right) $ \cite{SCATTERING} with $B_{0}=822$ \textrm{G} and $N=2\times
10^{6} $, $\omega _{x}/2\pi =\omega _{y}/2\pi =115\times \sqrt{25}$ \textrm{%
Hz}, $\omega _{z}/2\pi =1.1\times \sqrt{25+120\times B/1000}$ \textrm{Hz}.

The solid line in figure 2 shows the theoretical result based on the above
equation (\ref{Ketterle}). We see that our theory agrees well with the
experimental result of \cite{MIT-fermion} illustrated by $\square $ in
figure 2. Different from the analyses for the experiment by JILA group, we
do not consider the loss of molecules during the probing procedure because
in the experiment by MIT group for $^{6}Li$, the molecules is much more
stable comparing with the JILA group with $^{40}K$. We believe that both the
experimental data of JILA group \cite{JIN-fermion} and MIT group \cite%
{MIT-fermion} are reliable. The large difference of the condensate fraction
in \cite{JIN-fermion} and \cite{MIT-fermion} is due to different
experimental process.

\begin{figure}[tbp]
\includegraphics[width=0.8\linewidth,angle=270]{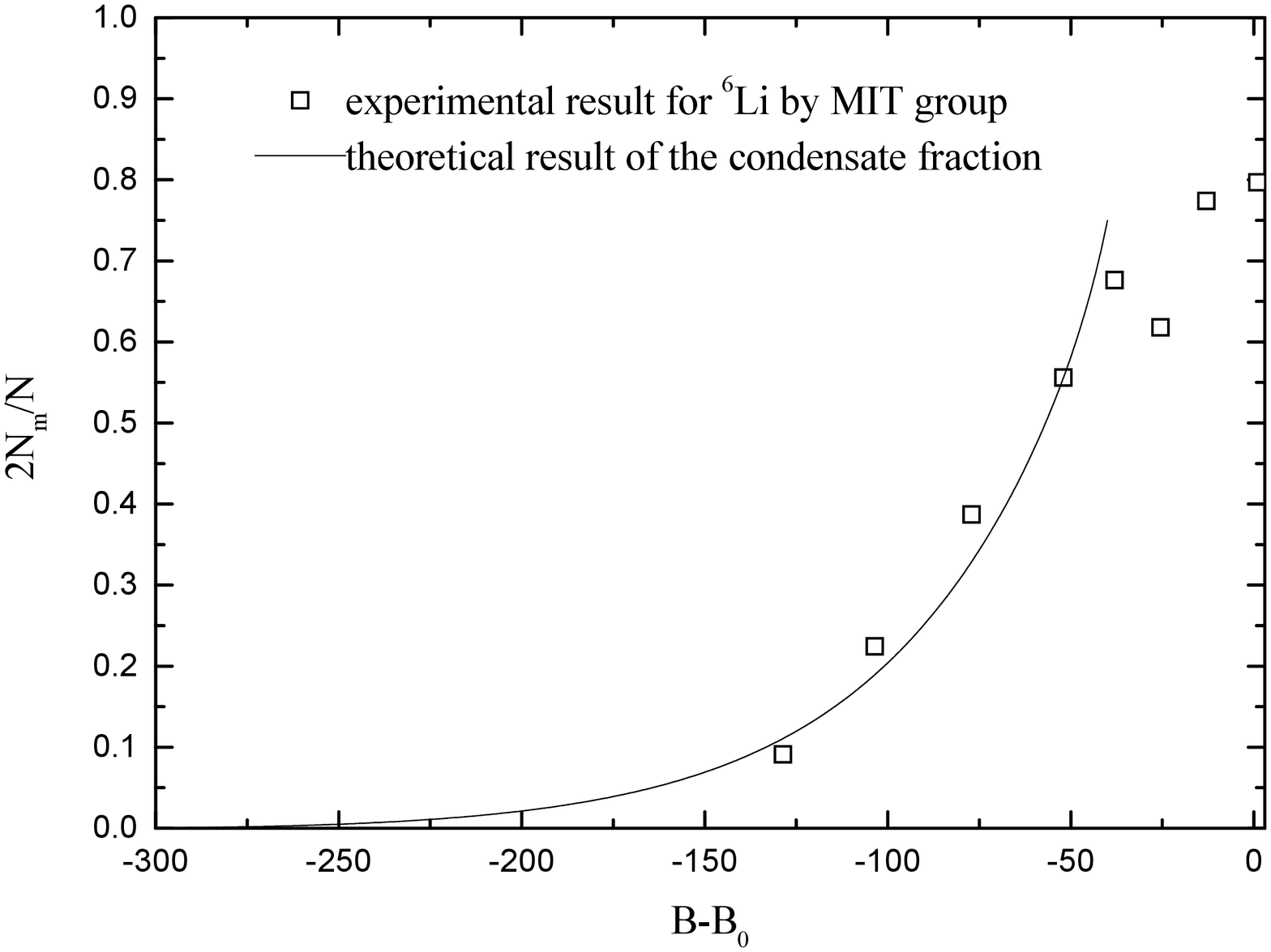}
\caption{The solid line shows the calculated condensate fraction
of molecules
at zero temperature for an experimental process of MIT group \cite%
{MIT-fermion} where the ultracold gases with condensate fraction
$0.56$ were initially prepared at the magnetic field $770$
\textrm{G} below the Feshbach resonant magnetic field $B_{0}$. We
see that the solid line based on the energy conservation given by
Eq. (\ref{Ketterle}) agrees well with the
experimental result shown by $\square $ with a hold time of $100$ \textrm{ms}%
.}
\end{figure}

In the above theoretical investigation for the experiment by MIT group, the
harmonic frequency $\omega _{z}$ is dependent on the magnetic field. In
fact, due to the $B$-dependent harmonic frequency $\omega _{z}$, there is an
effective work on the system by the magnetic field due to the continuous
changing of the ground state of the system. This effective work can be
omitted because $\Delta \omega _{z}/\omega _{z}<4\%$ between the regime of
magnetic field $770$ \textrm{G} and $700$ \textrm{G} investigated in the
experiment. In figure 2, the agreement between theoretical result and
experimental data shows clearly that the one can safely omit the effective
work done by the magnetic field.

For the condensate fraction discussed here, the term $N_{m}\varepsilon _{m}$
in the overall energy of the molecular gas $E_{m}\left( N_{m},B\right) $
plays a special role because it is negative and dependent strongly on the
magnetic field. Below the Feshbach resonance and with the decreasing of the
magnetic field, the term $\varepsilon _{m}$ decreases significantly. When
the loss of energy is omitted, to maintain the energy conservation, there
should be a decreasing of the number of molecules when lowering
adiabatically the magnetic field, which has been shown clearly in \cite%
{JIN-fermion} and \cite{MIT-fermion}. One should note that the overall
energy of the ultracold gases is investigated here for the case of weakly
interacting gases with $a/\overline{l}<1$ and the case of the divergent
scattering length. Thus the theoretical result for the condensate fraction
obtained here does not hold in the strongly interacting gases with $a/%
\overline{l}>1$, and this is the reason why we do not give the condensate
fraction for $a/\overline{l}>1$.

In summary, the simple theory developed here is used to interpret different
experiments by JILA group \cite{JIN-fermion} and MIT group \cite{MIT-fermion}
on the condensate fraction. Obviously, the condensate fraction discussed
here is dependent on the magnetic-field sweep process and initial condition
of an experiment. When the ultracold gases are prepared at the magnetic
field below $B_{0}$, after different evaporative cooling one can get
relatively arbitrary condensate fraction. One can check our theory further
by investigating the condensate fraction in experiments for different
magnetic-field sweep process and different initial condition of the system.
It will be an interesting future work to investigate the condensate fraction
of the fermionic atom pairs above the Feshbach resonant magnetic field based
on the assumption that the loss of energy can be omitted.

H. W. Xiong acknowledges very useful discussions with J. E. Thomas and C.
Chin.



\begin{thebibliography}{99}
\bibitem{SCATTERING} $a_{m}=170$ \textrm{nm} for $B=770$ \textrm{G} and $%
a_{m}=66$ \textrm{nm} for $B=715$ \textrm{G} measured in \cite{SALOMON} are
used to get this form of the scattering length for $^{6}Li$.
\end{thebibliography}
\end{document}